\documentclass[twocolumn,prl,amsmath,amssymb]{revtex4}
\usepackage{graphicx}
\begin{document}
\newcommand{\lsig}{\lambda_{\sigma}^{-1}}
\newcommand{\lzr}{\lambda_{0}^{-1}}
\newcommand{\Grho}{\Gamma_{\rho}^{-1}}
\newcommand{\Gzr}{\Gamma_{0}^{-1}}
%\draft
\title{ Is there a true Model-D critical dynamics?}
\author{Parongama Sen}
\affiliation{Department of Physics, University of Calcutta,
92 A.P. C. Road, Kolkata 700009, India.} 
\email{parongama@vsnl.net, paro@cubmb.ernet.in}
\author{Somendra M. Bhattacharjee}
\affiliation{Institute of Physics, Bhubaneswar 751 005, India}
\email{   somen@iopb.res.in }
%\maketitle
\begin{abstract}                                 
  We show that non-locality in the conservation of both the order
  parameter and a noncritical density (model D dynamics) leads to new
  fixed points for critical dynamics.  Depending upon the parameters
  characterizing the non-locality in the two fields, we find four
  regions: (i) model-A like where both the conservations are
  irrelevant (ii) model B-like with the conservation in the order
  parameter field relevant and the conservation in the coupling field
  irrelevant (iii) model C like where the conservation in the order
  parameter field is irrelevant but the conservation in the coupling
  field is relevant, and (iv) model D-like where both the
  conservations are relevant.  While the first three behaviours are
  already known in dynamical critical phenomena, the last one is a
  novel phenomena due entirely to the non-locality in the two fields.

\end{abstract}

\maketitle

%\begin{multicols}{2}

Based on the long time behaviour, the dynamics of different critical
systems are classified \cite{HH} in various dynamic universality
classes.  These classes are characterized by several factors like the
number of conserved field, non-vanishing Poisson brackets etc, in
addition to the usual quantities necessary for equilibrium. Among the
models which have no propagating modes, Model A is a system in which
there are no conserved fields, Model B is a system with a conserved
order parameter while Model C is the class of systems with a
non-conserved order parameter field coupled to a conservative field.
A fourth classification was also made - Model D, in which both the
order parameter field and the coupled field are conserved.  This
model, however, was shown to be equivalent to Model B as the coupling
to the field becomes irrelevant in the large length long time scale
limit.

The dynamical behaviour in systems with one or more conserved field
like Models B, C and D in the above classification is usually studied
by considering local conservation.  However, conservation is by no
means necessarily local and in fact, non-local conservation is a more
general condition.  A few recent studies on critical or near-critical
dynamics show that non-locality in the conserved field may change the
dynamical class, besides the obvious modifications in nonuniversal
details.  This kind of effect of non-locality in the conservation has
been noted in the study of critical dynamics of model B and C and in
the numerical study of the early time effect of a model believed to
belong to the class of model C and on the zero temperature dynamics
(phase-ordering kinetics or Ostwald ripening)
\cite{bray1,bray2,ps,satya,zheng,conti}.  Besides the theoretical
interest on its own right, one of the important applications of
nonlocal dynamics is in speeding up numerical simulations, prompting
implementation of various types of nonlocal moves\cite{moves}.  Hence
the need of a general classification of effects of nonlocality in
dynamics.

In a sense, model D is the most generalised class of dynamical
critical phenomena with dissipative dynamics.  However, with local
conservation, it is equivalent to Model B showing no novel features.
Motivated by the results of model C\cite{ps}, we investigate the
problem of non-local conservation in Model D in the most general way
by keeping both the conservations in the order parameter field and the
coupled field non-local.  One of our main results is to show the
emergence of a true model D behaviour by nonlocal dynamics, unlike the
local conservation case.

We introduce non-locality by using a non-local kernel
\cite{bray1,bray2,ps} in the chemical potential keeping the continuity
equation intact.  Using $\phi$ for the order parameter and $m$ for the
secondary coupling parameter, the Hamiltonian of model D is given
by
\begin{eqnarray}
\label{eq:6}
H = \int d^d{\bf x}[\frac{1} {2} r\phi^2({\bf x}) + \frac{1}{2}(\nabla
\phi ({\bf x}))^2
+\tilde u\phi^4({\bf x}) \nonumber \\
     + \gamma \phi^2({\bf x})m ({\bf x})+\frac{1}{2}C^{-1}
m^2 ({\bf x})],
\end{eqnarray}
where $\phi^2({\bf x}) = \sum_{i=1}^{n}\phi_i^2({\bf x}), \phi^4({\bf
  x}) = [\sum_{i=1}^{n}\phi_i^{2}({\bf x})]^2, r=0$ is the mean field
critical point, and $C>0$ ensures noncriticality of $m$.  There is a
short distance cutoff which in momentum space is an ultraviolet
cutoff $\Lambda$. We shall set $\Lambda=1$ in the calculation. The
above Hamiltonian, Eq. (\ref{eq:6}) is the same 
as in Model C\cite{ps,HH}.  Making $\gamma = 0$ effectively reduces
the above model to model A.

\renewcommand{\thefootnote}{\fnsymbol{footnote}}

The dynamics $^{\footnotemark[7]}$ \footnotetext[7]{ Additional torques for $n=3$ (Heisenberg model)  may
  lead to new dynamic universality  classes\cite{jdas}. 
 We do not consider such effects.} is given by the following equations:
%\begin{mathletters}
\begin{subequations}
\label{eq:1}
\begin{eqnarray}
  \label{eq:1a}
\left(
  \frac{1}{\Gamma_{\rho}k^{\rho}}+\frac{1}{\Gamma_0} \right )
\frac{\partial \phi_{\bf k}}{\partial t} &=& -\frac{\delta H}{\delta
  \phi_{\bf -k}}
  +\eta ({\bf k},t) \\
%\end{eqnarray}
%\begin{eqnarray}
  \label{eq:1b}
\left(
  \frac{1}{\lambda_{\sigma}k^{\sigma}}+\frac{1}{\lambda_0} \right )
\frac{\partial m_{\bf k}}{\partial t} &=&  -\frac{\delta H}{\delta
  m_{-\bf k}}   + \zeta ({\bf k},t) . 
\end{eqnarray}
%\end{mathletters}
\end{subequations}
The noises $\eta$ and $\zeta$ obey
\begin{subequations}
%\begin{mathletters}
\begin{eqnarray}
\langle \eta \rangle =0, \langle \zeta \rangle = 0, \hfill \\
\langle \eta ({\bf k},t) \eta ({\bf k}',t')\rangle =  2 (\frac
{1}{\Gamma_\rho k^\rho } +  
\frac {1}{\Gamma_0}) \delta({\bf k}+{\bf k}')\delta(t-t'),\\
\langle \zeta ({\bf k},t) \zeta ({\bf k}',t')\rangle  =  2 (\frac {1}{\lambda
  _\sigma k^\rho } +   
\frac {1}{\lambda_0}) \delta({\bf k}+{\bf k}')\delta(t-t').
\end{eqnarray}
%\end{mathletters}
\end{subequations}
Here, $k=\mid {\bf k}\mid$, and we have introduced two different
parameters $\rho$ and $\sigma$ to denote the non-localities in the
order parameter field and the coupling field respectively. 
The dissipative or the kinetic coefficients ($\Gamma_0$ and
$\lambda_0$) are denoted by the
subscript $0$ while the transport coefficients ($\Gamma_{\rho}$ and
$\lambda_{\sigma}$)  by the corresponding powers of $k$.
For the
local case, both $\rho$ and $\sigma$ assume a value equal to 2.
Results are known for some limiting values of $\rho$ and $\sigma$: a)
($\rho =0, \gamma = 0$) corresponds to model A \cite {HH}, b) ($\rho
=2, \gamma =0$) and ($\rho = 2, \sigma = 2 $) give model B like
behaviour \cite{HH}, c) any value of $\rho > 0$ and $\gamma = 0$ is a
non-local Model B \cite{bray1,satya}, and d) $\rho = 0$ with any
$\sigma > 0$ is non-local model C \cite{ps}.

A Hamiltonian of the type of Eq. \ref{eq:6} occurs near the critical
wings of a tricritical system or in critical systems with additional
constraints. Model D dynamics are relevant in such cases, for example
in three or four component mixtures or in spin-1 or Potts
models\cite{lawrie}. A spin-$\frac{1}{2}$ example, though a bit
unrealistic, would be  an antiferromagnetic system with
conserved magnetisation (as in model C \cite{ps2,zheng}) as well as
conserved staggered magnetisation.  The Kawasaki dynamics can be used
to conserve the magnetisation maintaining at the same time that the
exchanges between opposite spins have to be done in such a way that the
order parameter is unchanged. 

A momentum-shell renormalization group approach is used to study the
long distance long time behaviour. In this approach, small scale
fluctuations in space (between $\Lambda e^{-\delta l}$ and $\Lambda$
in momentum space) are integrated out, and the cutoff rescaled to
$\Lambda$. One then obtains an effective
Hamiltonian and  effective equations for the dynamics valid for
longer scales. The universal features are obtained from the fixed
points of the flow equations for the various parameters of the
problem.  As per the usual RG scheme, any parameter that vanishes
(grows) in the long scale limit is called an irrelevant (relevant)
variable. A fixed point is then characterized by the set of relevant
parameters.  In this scheme the four dynamics classes correspond to
the relevance and irrelevance of $\lambda_{\sigma}$ and
$\Gamma_{\rho}$ as mentioned in the abstract.

The statics of this system is the same as in Model C and the fixed
point values of the parameters $r, u$ and $\gamma^2 C$ are as given in ref
\cite{ps}.  The dynamical equations, on the other hand, take the
forms:
\begin{subequations}
%\begin{mathletters}
\begin{eqnarray}
  \label{eq:2}
  \frac{\partial \lsig}{\partial l} &=& ( -z+\sigma +
  \frac{\alpha}{\nu}) \lsig\\
  \label{eq:2b}
  \frac{\partial \lzr}{\partial l} &=& ( -z+\frac{\alpha}{\nu}) \lzr + 
  (n \gamma^2 K_d) Q\\
  \frac{\partial \Grho}{\partial l} &=& ( -z+\rho + 2 )\Grho\\
  \frac{\partial \Gzr}{\partial l} &=& ( -z+2) \Gzr + 
  (4 C\gamma^2 K_d) Q\frac{1}{1+\frac{Q}{CP}}
\end{eqnarray}
%\end{mathletters}
\end{subequations}
where $Q=\Grho+\Gzr$, $P=\lsig+\lzr$ and 
$K_d= 2^{1-d}\pi^{-d/2}\Gamma (d/2) $. 
The parameters $\lambda_\sigma, \lambda_0, \Gamma_\rho, \Gamma_0, \gamma$ 
and $C$ appearing in the above equations are all $l$ dependent.
To simplify notation, this $l$-dependence has been  suppressed. 

From (\ref{eq:2b})  we get
\begin{equation}
\label{eq:Gamma}
\frac {\Gamma_0}{\lambda_0} = \frac{(1+x)n\gamma^2K_d}{z-\alpha/\nu}
\end{equation}
 where $x = \frac{\Gamma_0}{\Gamma_\rho}$
is the dimensionless ratio of the two coefficients for the primary
order-parameter (note: $\Lambda=1$). This quantity $x$ satisfies
\begin{equation}
\label{eq:xflow}
\frac {\partial x}{\partial l} = (\rho -Y)x,\quad {\rm where}\quad Y =
\frac{2\alpha}{\nu n }\ \frac{1}{\mu^{-1} + (x +1)^{-1}}. 
\end{equation}
Here we have introduced the ratios of the transport and kinetic
coefficients of $m$ with the kinetic coefficient of $\phi$, namely,
\begin{equation*}
\mu_\sigma = \frac {\Gamma
  _0C}{\lambda_\sigma }, \quad 
~~\mu_0 = \frac {\Gamma _0C}{\lambda_0}
\quad {\rm and}~~ 
 \mu = \mu_\sigma + \mu_0. 
\end{equation*}
By using the fixed
point value of $C$ and Eq. (\ref{eq:Gamma}), $\mu_0$ can be rewritten
as 
\begin{equation*}
\mu_0 = \frac{(x+1)\alpha /\nu}{2(z-\alpha /\nu)} .
\end{equation*}

The flow equation for $\mu_\sigma $ is\\ 
\begin{equation}
\frac{\partial \mu_\sigma }{\partial l} = \mu_\sigma [\sigma  -2 +
\alpha/\nu - Y] . 
\end{equation}

We consider the different solutions of the equations for $x$ and
$\mu_\sigma$.  Corresponding to the relevance of the conserved
quantities, one gets different regions in the $\rho-\sigma$ plane with
different dynamic critical behaviour.  It should be mentioned here
that for $n \geq 4$, the coupling $\gamma$ scales to zero and one
effectively gets a non-local Model B.  This has already been
considered and shown to be model A-like when globally conserved
\cite{bray1}.  Our discussions are for the $n < 4$ region where the
coupling survives.

There are three fixed points of $x$: $x = 0$, $x = \infty$ and a
non-zero finite fixed point value for $x$ from Eq. (\ref{eq:xflow}):

\medskip

Case 1. $x = 0$ in the long length scale limit:\\

This corresponds to non-local Model C: the conservation of the order
parameter is irrelevant. The fixed points and their stabilities of
non-local model C, which were obtained in Ref. \cite{ps} from the
solutions of $\mu_\sigma$, now depends on the value of $\rho$. In
general, $x=0$ will be a stable fixed point as long as $\rho < Y$.  We
focus on the stability of the various regions for non-zero values of
$\rho$.

(a) $\mu_\sigma$ has a finite fixed point value and $z = \sigma +
\alpha/\nu$.  This is a stable solution in the region $(2/n)-1 > p >
-1$ (where $p = \frac{\sigma-2}{\alpha/\nu}$) for $\rho = 0$. For
non-zero $\rho$ this remains stable for $\rho < p+1$.

(b) $\mu_\sigma = \infty$ : this is valid for $ p > (2/n)-1$.  Here $z
= 2 +2 \alpha/n \nu$.  This solution $x = 0$, $\mu_\sigma = \infty$ is
stable as long as $\rho <2\alpha /n\nu$.

(c) For $\mu_\sigma = 0$, the stability of $x= 0$ is valid only for $
p < -1$ and $\rho < O(\epsilon^2)$.  Here $ z = 2 +O(\epsilon^2)$ and
the behaviour is model A like.  Any non-zero value of $\rho$ thus
destroys this model A like region. Therefore, a region where both the
conservations are irrelevant is restricted strictly to $\rho = 0$.

\medskip

Case 2. A finite non-zero value of $x$ in the long length scale limit:
This occurs when $\rho = Y$.  In order that $\Gamma_0$ reaches a
finite fixed point value, we must have $ z = 2 +\rho$. In this case,
$\mu_\sigma$ again has the following
fixed points:\\
(a) $\mu_\sigma = \infty$ for $ \sigma -2 +\alpha/\nu > \rho$\\
(b) $\mu_\sigma = 0 $ for $ \sigma -2 +\alpha/\nu < \rho$\\
(c) $\mu_\sigma$  nonzero finite   for $ \sigma -2 +\alpha/\nu = \rho$\\

Physically a finite valued fixed point of $x$ means that the
conservation in the order parameter $\phi$ is relevant.  If at the same time
$\mu_\sigma$ attains a non-zero fixed point value, the conservation of
the coupling field $m$ is also relevant.  This 
occurs if the fixed points (a) and (c) are stable.

For case (a), $x = \frac {\rho - 2\alpha /\nu n}{2\alpha /\nu n}$ and
for non-negative value of $x$, $\rho > 2\alpha /n \nu$.  Hence
in between $\rho = 2\alpha/n\nu$ and $\rho = \sigma -2 +\alpha/\nu$, a
stable region is obtained where both the conservations are relevant.

For case (b), $x +1 = \frac {\rho(2z-\alpha/\nu)}{2\alpha^2/n \nu^2}$.
This has a finite valued solution for $x$ only for $\rho \sim
O(\epsilon^2)$.  Hence for $\rho > \sigma -2 +\alpha/\nu$, $x
\rightarrow \infty$ is the only solution for $\mu = 0$. It is a
conventional model B fixed point.

Case (c): Here $z = \sigma +\alpha /\nu$ and in principle along the
entire line $\sigma -2 +\alpha/\nu = \rho$ a solution exists where
both the conservations are relevant.  However, there is a restriction
on the value of $\sigma$ from the condition of existence of a non-zero
finite value of $\mu_\sigma$.  We find that
\begin{equation}
\mu_\sigma +\mu_0 = \frac {(\sigma -2 +\alpha/\nu)(x+1)}{2-\sigma
  -\alpha/\nu + 
\frac{2\alpha}{n\nu}(x+1)}.
\end{equation}
Hence with $\sigma = 2 +p\alpha/\nu$; we get the condition $(1+p) <
2(x + 1)/n$.  Since $x$ is also non-zero, this implies that this
solution with both $x$ and $\mu_\sigma$ finite is valid for $p+1 =
\rho/(\alpha/\nu)$ with at least $p < (2/n) - 1$.

In the rest of the $\rho-\sigma$ plane, the only solution is $x =
\infty$ and $1/\lambda_\sigma =0$ for which only the conservation of the
order parameter is relevant. Here $z = \rho +2$.

% figure goes here
\begin{figure}
\includegraphics[clip,width=8cm]{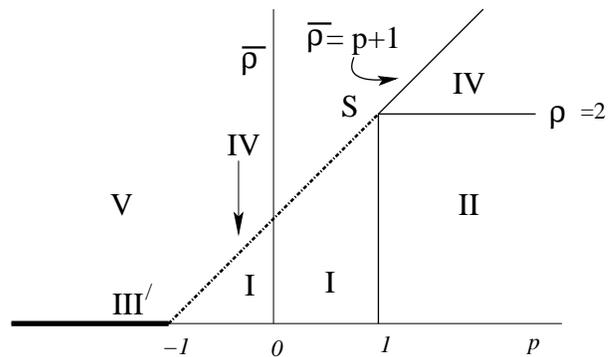}

\caption{The different regions in the non-local Model D
  with $n=1$ is shown in the rescaled $\bar{\rho} - p$ plane where
  $\bar{\rho}=\rho/( \alpha/\nu)$ and $p = \frac {(\sigma -2 )}{ \alpha
    /\nu}$.  Regions I and II are model C like with $ z= \sigma + \alpha
  /\nu$ and $z=2+2\alpha/\nu$ respectively.  Region III' is model A
  like (as in model C of Ref. 4) with $z = 2 + O(\epsilon
  ^2)$. Reg IV is the "Model D" region, 
  $z = \rho +2$; Reg V is Model B like, $z = \rho +2$.  The line $p+1
  =\rho$ is also model D-like upto point S (shown by the dashed
  line).  There are no discontinuities of $z$ along any boundary
  except at the boundary between III' and V.}
\label{fig:1}
\end{figure}

All the above possibilities of different dynamic behaviours are
summarized in Fig. 1 in the $\rho-\sigma$ plane (for $n=1$). The
possibility of seeing a region where both the conservation conditions
are relevant raises new issues like early time effect and boundary or
surface effects\cite{early} in the new regime.  These remain to be
studied.

To summarize, Model D can be viewed as the most general dynamic model
with purely dissipative dynamics and we indeed obtain all the four
types of behaviours as soon as non-locality in the conservations (as in
Eq. (\ref{eq:1})) are introduced.  Since our general model subsumes
the previously studied nonlocal model C and local model D, it is
natural to expect A-like, B-like and C-like regimes.   The
model A like region, however,  turns out to be unstable for any finite value of
$\rho$.  The model C like regions are stable for small nonzero values of 
$\rho$
but ultimately disappear for larger values of $\rho$.  The
most non-trivial result, for general values of $\rho$ and $\sigma$, is
the appearance (Region IV in Fig. 1) of a region with a new dynamical
behaviour where both the conservations are relevant. This we call a
{\it true model D like region} -- a consequence of nonlocal conservation laws.

Acknowledgments: PS acknowledges financial support from DST project
SP/S2/M-11/99.

%\begin{references}

%\begin{figure}
%\includegraphics{modeld_fig1.eps}

%\caption{The different regions in the non-local Model D
%  with $n=1$ is shown in the rescaled $\bar{\rho} - p$ plane where
%  $\bar{\rho}=\rho/( \alpha/\nu)$ and $p = \frac {(\sigma -2 )}{ \alpha
%    /\nu}$.  Regions I and II are model C like with $ z= \sigma + \alpha
%  /\nu$ and $z=2+2\alpha/\nu$ respectively.  Region III' is model A
%  like (as in model C of Ref. 4) with $z = 2 + O(\epsilon
%  ^2)$. Reg IV is the "Model D" region, 
%  $z = \rho +2$; Reg V is Model B like, $z = \rho +2$.  The line $p+1
%  =\rho$ is also model D-like upto point S (shown by the dashed
%  line).  There are no discontinuities of $z$ along any boundary
%  except at the boundary between III' and V.}
%\label{fig:1}
%\end{figure}

\end{document}